\documentclass[onecolumn,groupedaddress,showkeys]{ws-ijmpc}
\usepackage{float}
\usepackage{graphicx}
\usepackage{amssymb}
\usepackage{amsmath}
\usepackage{caption}
\usepackage{subcaption}
\usepackage{xcolor}
\usepackage[super]{cite} 
\usepackage[breaklinks]{hyperref}
\hypersetup{colorlinks,urlcolor=black,citecolor=black,linkcolor=black,filecolor=black}
\usepackage{breakurl}
\usepackage{anysize}
\marginsize{3cm}{3cm}{3cm}{3cm}
\begin{document}
	
\markboth{Sourav Chowdhury , Suparna Roychowdhury \& Indranath Chaudhuri}
{A robust prediction from a minimal model of COVID-19 - Can we avoid the third wave?}

\catchline{}{}{}{}{}

\title{{A robust prediction from a minimal model of COVID-19 - Can we avoid the third wave?}
}

\author{Sourav Chowdhury\thanks{email: chowdhury95sourav@gmail.com}\qquad Suparna Roychowdhury\thanks{email: suparna@sxccal.edu}\qquad Indranath Chaudhuri\thanks{email: indranath@sxccal.edu}}

\address{Department of Physics, St. Xavier's College (Autonomous)\\
	30 Mother Teresa Sarani, Kolkata-700016, West Bengal, India}

\maketitle

\begin{history}
	\received{Day Month Year}
	\revised{Day Month Year}
\end{history}

\begin{abstract}		
	COVID-19 pandemic is one of the major disasters that humanity has ever faced. In this paper, we try to model the effect of vaccination in controlling the pandemic, particularly in context to the third wave which is predicted to hit globally. Here we have modified the SEIRD model by introducing a vaccination term. One of our main assumptions is that the infection rate ($\beta(t)$) is oscillatory. This oscillatory nature has been discussed earlier in literature with reference to the seasonality of epidemics. However, in our case we invoke this nature of the infection rate ($\beta(t)$) to model the cyclical behavior of the COVID-19 pandemic within a short period. This study focuses on a minimalistic approach where we have logically deduced that the infection rate ($\beta(t)$) and the vaccination rate ($\lambda$) are the most important parameters while the other parameters can be assumed to be constants throughout the simulation. Finally, we have studied the rich interplay between the infection rate ($\beta(t)$) and the vaccination rate ($\lambda$) on the infectious cases of COVID-19 and made some robust conclusions regarding the global behavior of this pandemic in near future. 
	
	\keywords{COVID-19; SEIRD model; minimal model; intervention; periodic infection rate; vaccination; third wave.}
\end{abstract}

\ccode{PACS Nos.:}

\section{Introduction}
	COVID-19 pandemic is a curse on humanity. This disease has taken many lives but also has hampered the social and economic developments of countries around the globe. Globally, 221,326,243 people have been infected and 4,579,475 people have died from this pandemic as on 04/09/2021. India has badly suffered from the pandemic. 32,987,615 people have been infected and 440,567 people have died already in India until 04/09/2021 \cite{worldometers}. Most of the countries already have faced multiple (two or three) waves of this disease. Also, India has faced a large second wave of this disease where the period of it is very small. So, this wave of COVID-19 was rose very fast and almost becomes uncontrollable. So, the government has no chance but to apply complete lockdown. We are currently trying to overcome the devastating effect of this. Given the fact that the virus has shown mutations and certain countries are facing a new spike of COVID-19, there is a high possibility that India will be hit by a fresh onset of the third wave \cite{third_wave_model_stat,third_wave_model}. This is a serious concern for us since we are already very compromised in terms of our economy and health facilities after the devastation by the second wave. The only long-term way to control this pandemic is adequate vaccination. Several authors have addressed the importance of vaccination from a different view points in context to COVID-19 pandemic \cite{chaos_vac,chaos_vac_osc,covid_vac_india,covid_vac_acc_india,NPI_vs_vac,vac_resistant,phase_vac, CA_vac}. There are various studies which are related to different aspects of COVID-19 pandemic and other epidemics around the globe \cite{HIT, immunization, spacetime_covid, fractal_covid, percolation_covid, multistage_CA_GA}. Also, the socio-economic factors of a country can affect the dynamics of an epidemic or a pandemic.\cite{sociophysics_book, covid_slum}  In this paper, we have tried to find the effect of vaccination on the third wave by mathematical modeling from a different angle. 
	
	To model the effect of vaccination on this cyclical behavior of COVID-19 we are using SEIRD (Susceptible-Exposed-Infectious-Recovered-Dead) model \cite{basic_SIR, keeling2011modeling,SEIRD_covid,seird_vac_forced}.  Our main motivation to study is to look at the possibility of the resurgence of the third wave by studying the interplay between the infection rate and the vaccination rate. In this model, one of the important assumptions is that the infection rate ($\beta (t)$) is oscillatory. Previously, various authors have used the oscillatory behavior of the infection rate ($\beta (t)$) to model the seasonal recurrence of many other epidemics and diseases \cite{chaos_vac,period_doubling,seasonality,chaos_vac_osc,seasonality_lit_rev}. Here, in our study, the oscillatory nature of the infection rate ($\beta (t)$) has been modeled for COVID-19 from a different perspective, the details of which are discussed later in the paper. 
	
	We have first studied the evolution of COVID-19 without any vaccination. Next, we have included a term which represents vaccination in this model. The two main free parameters which models the vaccination term are the vaccination rate ($\lambda$) and the time at which vaccination is started ($\tau$) after the onset of the pandemic. Thus we see that in addition to the infection rate ($\beta(t)$), we also have another important parameter - vaccination rate ($\lambda$), which influences the outcome of this model. Here our main assumption about the vaccination term is that a person will not be immune to this disease immediately after vaccination. Thus a delay time ($\sim$ time to become immune), which is represented by $\frac{1}{k}$ has been included in the equations. Next it was deduced through some careful arguments that all the other parameters can be fixed to some constant values since they have minimal effect on the evolution of the pandemic. So, we can focus our attention on the interaction between the infection rate ($\beta(t)$) and the vaccination rate ($\lambda$) primarily to determine the spread of COVID-19.
	
	In this minimal model, we also aim to investigate the value of the vaccination rate ($\lambda$)) and the vaccination starting time ($\tau$) which will suppress the third wave of this pandemic and keep its spread under control. We also try to explore a range of values of $\lambda$ and $\tau$ and tune them appropriately for different assumptions of the other parameters of the model to observe there effect on the diminution of the spread of this pandemic beyond the second wave.
	
	Finally, the arrangement of our paper is as follows: Section~\ref{Mathematical Model} consists of the detailed description of the mathematical model and its parameters. Section~\ref{Vaccination approximation} gives an approximate insight about the vaccination rate of our model. In Section~\ref{sim_section} we present the analysis and simulations of our model considering different cases and finally we end with a concluding section that summarizes the important features and results of our model in context to the future behavior of the COVID-19 pandemic.
\section{Mathematical Model}\label{Mathematical Model}
\begin{figure}[H]
		\centering
		\includegraphics[scale=0.55]{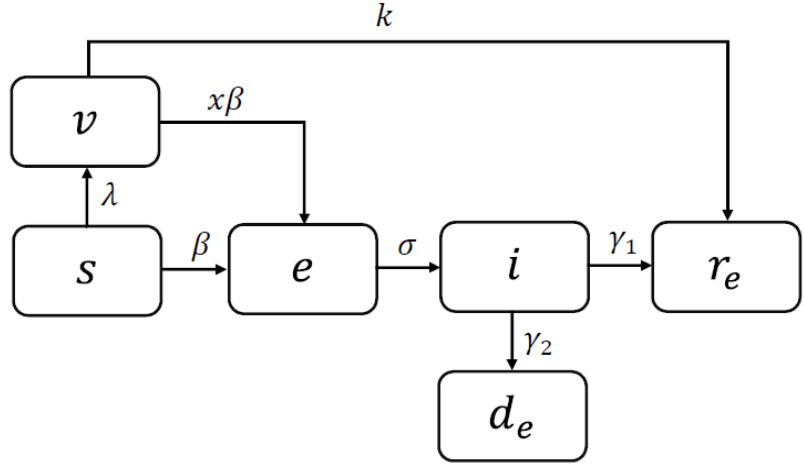}
		\caption{{Block diagram of our model.}\label{block_diagram}}
	\end{figure}

	In this paper, we have modified the SEIRD 
	(Susceptible-Exposed-Infected-Recovered-Dead) model by introducing a vaccination term. The block diagram of our model is shown in Fig.~\ref{block_diagram}. The main assumptions of our model are: 
	\begin{itemize}
		\item Vaccination is started with a delay $\tau$ from the initial time $t=0$.
		
		\item After being vaccinated, a person moves to a temporary vaccination compartment which is represented by $v(t)$.
		
		\item Fraction of population who are in the temporary vaccinated compartment ($v(t)$) are assumed to have not developed immunity against COVID-19 yet. They develop immunity with a delay of $\frac{1}{k}$ after the vaccination.
		
		\item It is assumed that vaccinated fraction of population are prone to being infected by this disease with a lesser probability, given by $x$.
		
		\item Thus a person can move from the $v(t)$ compartment either by getting infected or when they are fully immune to this disease. When infected, the person will move to the exposed compartment ($e(t)$). However, if they are fully immune to this disease after the vaccination, they will move to the recovered compartment ($r_{e}(t)$). 
	\end{itemize}

	 Now, we describe the modified model equations which are given below: 
	\begin{equation}
		\frac{ds}{dt}=-\beta si-\lambda s\label{s_eq}
	\end{equation}	

	\begin{equation}
		\frac{de}{dt}=\beta si-\sigma e+x\beta vi
	\end{equation}

	\begin{equation}
		\frac{di}{dt}=\sigma e-(\gamma_{1}+\gamma_{2})i
	\end{equation}

	\begin{equation}
		\frac{dr_{e}}{dt}=\gamma_{1}i+kv
	\end{equation}

	\begin{equation}
		\frac{dd_{e}}{dt}=\gamma_{2}i
	\end{equation}
	
	\begin{equation}
		\frac{dv}{dt}=\lambda s-x\beta vi-kv
	\end{equation}
	Here all the variables are fractions of the total population. So, we can write, 
	\begin{equation}
		s+e+i+r_{e}+d_{e}+v=1.\label{tot_pop}
	\end{equation}
	Descriptions of all variables and parameters in our model are given in the following two tables, Tab.~\ref{des_var_tab} and Tab.~\ref{des_par_tab}.
	\begin{table}[H]
		\tbl{Descriptions of the all variables in the model.}
		{\begin{tabular}{| c | l |} 
			\hline
			Variables & Description of the variables \\
			\hline
			$s$ & Fraction of susceptible population.\\
			$e$ & Fraction of exposed population.\\
			$i$ & Fraction of infectious (person who can spread disease) population.\\
			$r_{e}$ & Fraction of recovered population.\\
			$d_{e}$ & Fraction of dead population.\\
			$v$ & Fraction of population who has been vaccinated but not immune from the disease.\\ 
			\hline
		\end{tabular}
		\label{des_var_tab}}
	\end{table}
	
	\begin{table}[H]
		\tbl{Descriptions of the all parameters in the model.}
		{\begin{tabular}{| c | l |} 
			\hline
			Parameters & Description of the parameters \\
			\hline
			$\beta$ & infection rate.\\
			$\frac{1}{\sigma}$ & Latency period of the disease.\\
			$\gamma_{1}$ & Recovered rate.\\
			$\gamma_{2}$ & Death rate.\\
			$\lambda$ & Vaccination rate.\\
			$\frac{1}{k}$ & Time require to develop immunity due to vaccination.\\
			$x$ & Susceptibility of infection.\\
			$\tau$ & Delay in vaccination.\\
			\hline
		\end{tabular}
		\label{des_par_tab}}
	\end{table}
	In Eq.~\ref{s_eq}, $-\lambda s$ term denotes the decrement of the fraction of the susceptible population per unit time due to vaccination. So, $\lambda s$ denotes the increment of the fraction of the vaccinated population per unit time. Let, $v_{tot}(t)$ denote the fraction of the total vaccinated population at time $t$. This fraction includes all vaccinated people irrespective of their immunity. So, we can write,
	\begin{equation}
		\frac{dv_{tot}}{dt}=\lambda s\label{vtot_eq}
	\end{equation}
	We will revisit the above equation (Eq.~\ref{vtot_eq}) in section~\ref{Vaccination approximation} to gain some simple insights into this model.
	\subsection{Model of the infection rate, $\beta(t)$}
		Here we have assumed that the infection rate ($\beta(t)$) has an oscillatory nature. As we have mentioned before, the oscillatory behavior of the infection rate ($\beta(t)$) has been used in different studies of epidemics earlier from the perspective of seasonal re-occurrence of these diseases. \cite{chaos_vac,period_doubling,seasonality,chaos_vac_osc,seasonality_lit_rev}. Inspired from this idea, we have introduced a similar form of the infection rate ($\beta(t)$) in our study of the COVID-19 pandemic. This periodic nature has been incorporated here to model the aspect of human intervention in the form of social distancing, masking, and lockdown.
		
\begin{figure}[H]
			\centering
			\includegraphics[scale=0.34]{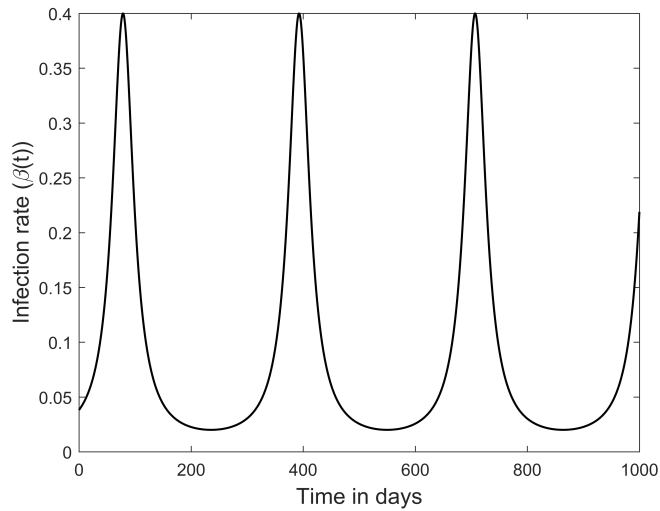}
			\caption{{Variation of $\beta(t)$ for $\beta_{P}$=0.4 $day^{-1}$, $\beta_{min}$=0.02 $day^{-1}$, and $\omega$=0.02 $day^{-1}$.}\label{var_beta}}
		\end{figure}

		It is observed that when the infection rate was high, governments had put restrictions to constrain or decrease the infection rate. Also, people became more cautious and maintained all appropriate measures like social distancing and masking. Thus the infection rate decreased. However, when the infection rate became low, the interventions were lifted (due to economic and social reasons) and  people became casual. In such situations, it was seen that the infection rate again started to increase. Thus the infection rate ($\beta(t)$) is seen to have such periodic cycles of increase and decrease. So, we can say that when $\beta(t)$ is maximum, human intervention is minimum and vice-versa. Here, we also note that the high infection rate does not imply a large number of infected cases. It only means that the chance of a susceptible person being infected by an infectious person is higher. Thus it can be concluded that a suitable function for the infection rate ($\beta(t)$), which has a sharp peak, can model the rapid rise and fall of the infection rate due to the variations in the intervention processes effectively. Such an oscillatory and periodic behavior in the infection rate ($\beta(t)$) can be emulated by a mathematical model which is given below:
		
		\begin{equation}
			\beta(t)=\frac{1}{a-b\sin(\omega t)} \label{beta_fun}
		\end{equation}
		where $\omega$ is the frequency of the infection rate ($\beta(t)$). Hence, the time period of the infection rate ($\beta(t)$) is $T=\frac{2\pi}{\omega}$. If $a>b$,  from Eq.~\ref{beta_fun} we see that $\beta_{max}=\beta_{P}=\frac{1}{a-b}$ and $\beta_{min}=\frac{1}{a+b}$. Thus in this case, the infection rate ($\beta (t)$) is completely defined by $\beta_{P}$, $\beta_{min}$, and $\omega$. In our study, we have assumed that $\beta_{min}\sim0.1\beta_{P}$ and we have chosen a fixed value of $\beta_{min}$. However, $a<b$ is inconsistent for our study since $\beta(t)$ cannot be negative. Fig.~\ref{var_beta} shows a representative behavior of the infection rate ($\beta(t)$) for a particular value of the $\beta_{P}$, $\beta_{min}$, and $\omega$.
		
		So, peak of the infection rate ($\beta_{P}$) and the frequency of the infection rate ($\omega$) are the most important quantities for the variation of the infection rate ($\beta (t)$) (as, $\beta_{min}$ is assumed to be fixed in our study). In section~\ref{sim_section} we have varied $\beta_{P}$ and $\omega$ to find the variations of the fraction of the infectious cases ($i(t)$) for both cases, with and without vaccination. 

		We have also studied the variation of the fraction of the infectious cases ($i(t)$) with different forms of $\beta(t)$ like sinusoidal function, damped oscillatory function, and linear superposition of two sinusoidal functions. To conclude, the form of the infection rate ($\beta(t)$), as given in Eq.~\ref{beta_fun}, was seen to be most suitable to model the behavior of the infectious cases for the COVID-19 pandemic.
	\subsection{Initial conditions and important parameters - A discussion}
		Initially, in a population, there are no recovered, dead or exposed person when a disease had not spread at all. So, $r_{e}(0)=0$, $d_{e}(0)=0$, and $e(0)=0$. We can also take $v(0)=0$, as it is assumed that vaccination started with a delay $\tau$. So, initially, there exists only two non-zero fraction of populations $s$ and $i$. Thus these two are bound by the relation $s(0)+i(0)=1$.
		
		In this model $\frac{1}{\sigma}$ and $\frac{1}{\gamma_{1}+\gamma_{2}}$ represent the mean latency period and the mean infectious period respectively. So, for COVID-19, we can assume some fixed values for them. Also, $\frac{1}{k}$ represents the mean time to develop immunity after vaccination. This implies that we can also fix $\frac{1}{k}$ to a certain value which is relevant for COVID-19. Thus, $\sigma$, $\gamma_{1}$, $\gamma_{2}$, and $k$ are those among all the parameters of our model to be fixed through out the entire work.
	\subsection{Significance of the parameter $x$}
		We have assumed that immediately after vaccination a person will not be immune to COVID-19, as has been observed and predicted in literature, a person takes some time to become immune from the COVID-19 after vaccination. In this period a person can be infected by this disease. Here, we have also assumed that these people are less susceptible to the disease than a susceptible person. Here, $x$ represents the susceptibility of these people to be infected from COVID-19. Thus $x<1$.
		
\section{Vaccination approximation}\label{Vaccination approximation}
	In this section, we try to simplify our model using certain approximations to qualitatively interpret the significance of the vaccination rate. The fraction of the total vaccinated population ($v_{tot}(t)$) depends on the susceptible population which is coupled with the whole model. So, it is very difficult to determine the exact time dependence of the fraction of the total vaccinated population ($v_{tot}(t)$). However, it is possible to find an approximate time dependence of this quantity using a simple analysis. The initial assumption is that at any time $t$, $i\ll1$, refer to the section~\ref{good_apprx}. So, Eq.~\ref{s_eq} becomes, 
	\begin{equation}
		\frac{ds}{dt}\approx -\lambda s\label{apprx_s_eq}
	\end{equation}	
	Let at time $\tau$ the vaccination is started. Hence, solving Eq.~\ref{apprx_s_eq} we can write, 
	\begin{equation}
		s(t)=s(\tau)e^{-\lambda (t-\tau)},\qquad t\ge\tau \label{s_form}
	\end{equation}	
	Substituting Eq.~\ref{s_form} in Eq.~\ref{vtot_eq} and assuming $v_{tot}(\tau)=0$,
	\begin{equation}
		v_{tot}(t)=s(\tau)\left(1-e^{-\lambda (t-\tau)}\right),\qquad t\ge\tau \label{vtot_form}
	\end{equation}	
	
	Our second assumption is that the fraction of the total infected cases at the vaccination starting time ($\tau$), $i_{tot}(\tau)\ll1$. Thus $s(\tau)\approx 1$. Including this approximation in Eq.~\ref{vtot_form} we can write,
	\begin{equation}
		v_{tot}(t)\approx\left(1-e^{-\lambda (t-\tau)}\right),\qquad t\ge\tau \label{vtot_form_approx}
	\end{equation}
	If $(t-\tau)=\frac{1}{\lambda}$,
	\begin{equation}
		v_{tot}=\left(1-\frac{1}{e}\right)=0.6321 \label{vtot_val}
	\end{equation}
	This implies that it will take $\frac{1}{\lambda}$ number of days (after the start of vaccination) to vaccinate 63.2\% (which comes from the exponential form of the $v_{tot}(t)$, Eq.~\ref{vtot_form_approx} and Eq.~\ref{vtot_val}) of the susceptible population or the total population (in this case susceptible and the total population is approximately the same as $s(\tau)\approx 1$).
	
	\begin{figure}[H]
		\begin{subfigure}{.49\textwidth}
			\centering
			\includegraphics[scale=0.26]{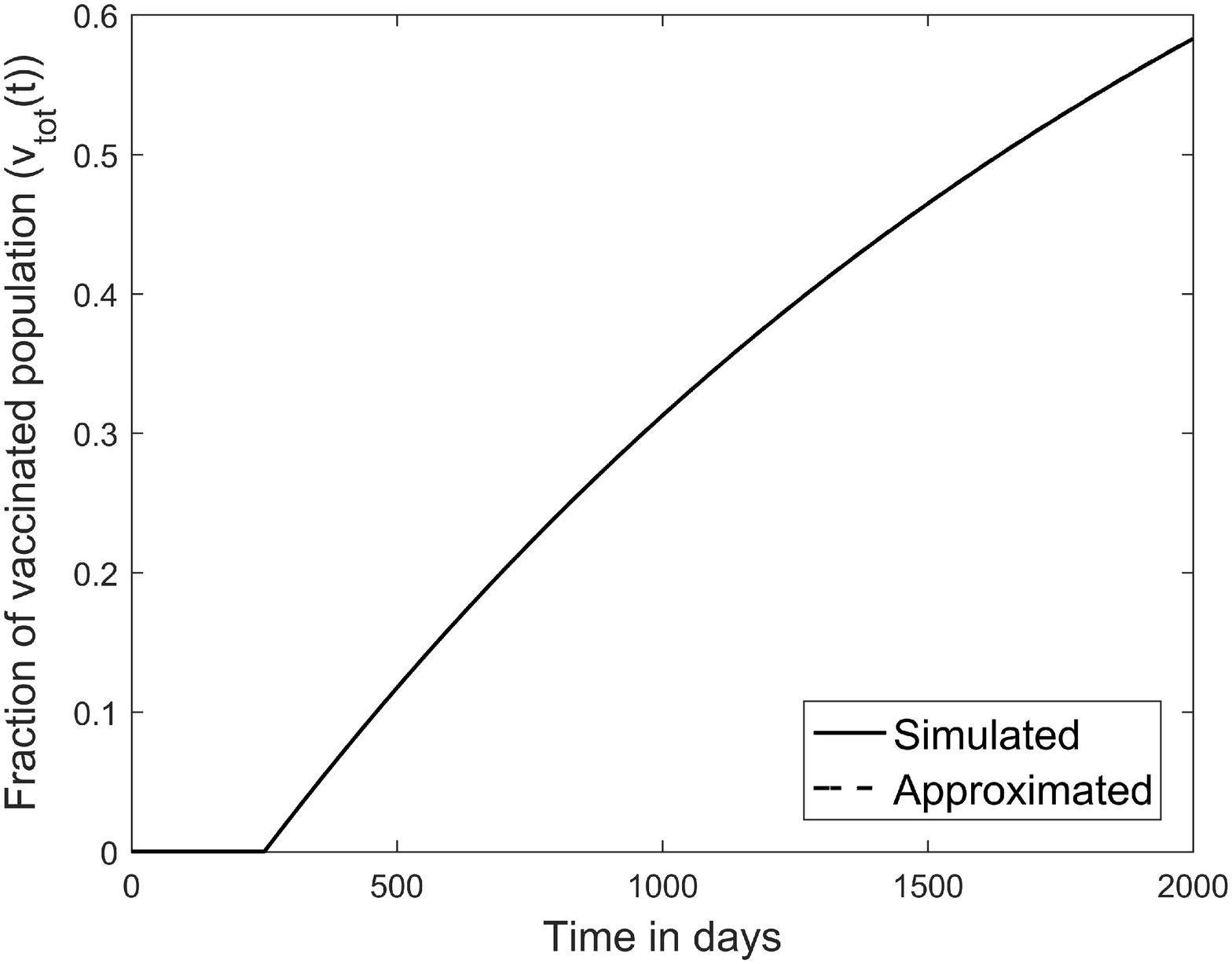}
			\caption{}
		\end{subfigure}
		\begin{subfigure}{.49\textwidth}
			\centering
			\includegraphics[scale=0.26]{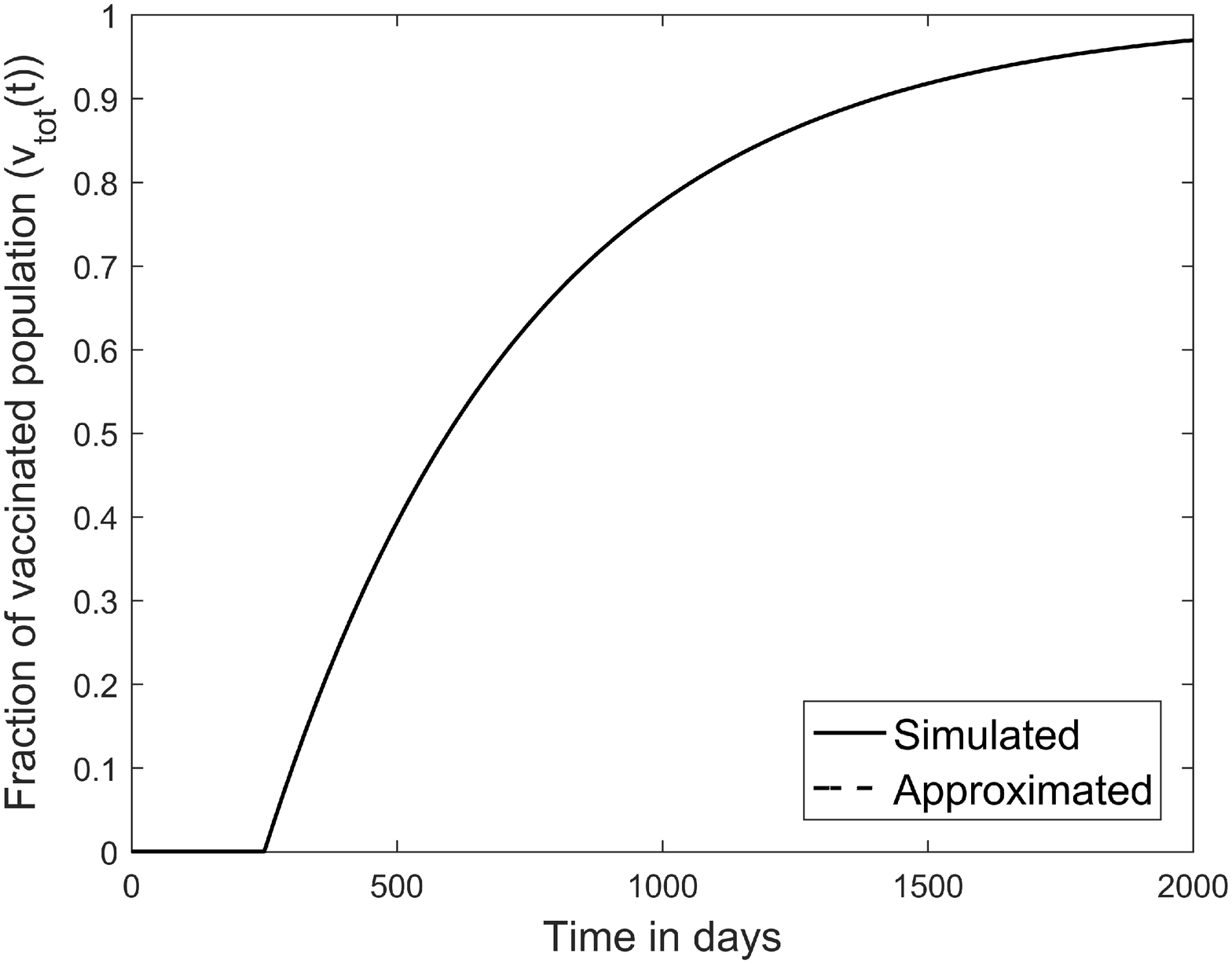}
			\caption{}
		\end{subfigure}
		\caption{{Comparison of the simulated and the approximated fraction of the vaccinated population ($v_{tot}(t)$) for different values of the vaccination rate ($\lambda$). (a): $\lambda$=0.0005 $day^{-1}$. (b): $\lambda$=0.002 $day^{-1}$.}\label{comp_vtot}}	
	\end{figure}

	In Fig.~\ref{comp_vtot} we have compared the simulated and approximated fraction of the total vaccinated population ($v_{tot}(t)$) as a function of time, $t$. This figure shows that whatever be the value of the vaccination rate ($\lambda$), simulated and approximated curves follow each other precisely. The initial conditions and the parameter values which have been used to generate these plots are discussed in section~\ref{sim_section}.
	
	We note that the growth of the fraction of the total vaccinated population ($v_{tot}(t)$) emulates the time dependence of the charging of a capacitor. If the charging medium is removed, the capacitor starts to discharge exponentially with time. Hence it is necessary to charge a capacitor after a specific time interval to maintain its charge. An analogous behavior (like a discharging capacitor) may be observed in the fraction of the total vaccinated population ($v_{tot}(t)$) if the vaccination is completely stopped.  This can happen because a vaccinated person loses his/her immunity with time. Thus the vaccination campaign has to be run in specific intervals to boost the immunity of the population with time. This aspect could form the basis of interesting studies in future.
	\subsection{Is this a good approximation?}\label{good_apprx}
		\begin{itemize}
			\item According to our model a recovered person cannot be infected again. So, the vaccination will be a determining factor to prevent the rise of the infectious cases if at a time $\tau$ there exists a large susceptible people. This implies, $s(\tau)\approx 1$ or, $i_{tot}(\tau)\approx 0$. This fact is true for any disease and its spread.
			
			\item At any time during COVID-19 pandemic, the total number of infected cases and the active cases are still much lower compared to the total population of the respective countries. We have shown this in Tab.~\ref{real_dat_tab}, where the data have been taken from \cite{worldometers}. This implies that $i_{tot}(\tau)\ll 1$ and $i(t)\ll1$. Thus the data has well qualifies these approximations in the case of COVID-19 pandemic. Note that the last column indicates the fraction of the total infected cases at the current time as mentioned the date. Whereas, the approximations have been made considering the fraction of the total infected cases at the starting of vaccination.
		\end{itemize}
	
		\begin{table}[H]
			\tbl{Table of the active cases, total infected cases and the respective fractions (with respect to the total population) of the five most infected countries for COVID-19 pandemic as on 04/09/2021.}
			{\begin{tabular}{| c | c | c | c | c | c |} 
				\hline
				Country & Population &Active cases & Total infected cases  & Fraction of & Fraction of\\
				&&&& active cases & total infected cases\\
				\hline
				USA & 333,278,410 & 8,772,892 & 40,703,674 & 0.02632 & 0.12213\\
				India & 1,395,901,836 & 405,650	 & 32,945,907 & 0.00029 &  0.02360\\
				Brazil & 214,330,637 & 453,105 & 20,856,060 & 0.00211 & 0.09731\\
				Russia & 146,007,887 & 552,825 & 6,975,174 & 0.00379 & 0.04777\\
				UK & 68,304,162 & 1,210,000 & 6,904,969 & 0.01771 & 0.10109\\
				\hline
			\end{tabular}
			\label{real_dat_tab}}
		\end{table}	
	
\section{Simulation of our model}\label{sim_section}
	In this section, we will first simulate our model without vaccination and then with vaccination. The values of the parameters that are used in our model is shown in Tab.~\ref{val_par_tab} \cite{latency_period,recov_time}.
	\begin{table}[H]
		\tbl{Values of the fixed parameters in the model.}
		{\begin{tabular}{| c | c |} 
			\hline
			Parameter & Value of the parameter \\
			\hline
			$\beta_{min}$ & 0.02 $day^{-1}$.\\
			$\frac{1}{\sigma}$ & 5.1 $days$.\\
			$\frac{1}{\gamma_{1}+\gamma_{2}}$ & 21 $days$.\\
			$\frac{1}{k}$ & 30 $days$.\\
			$x$ & 0.1.\\
			$\tau$ & 250 $days$.\\
			\hline
		\end{tabular}
		\label{val_par_tab}}
	\end{table}

	\begin{table}[H]
		\tbl{Initial values of the variables of our model.}
		{\begin{tabular}{| c | c |} 
			\hline
			Variable & Initial value of the variable \\
			\hline
			$e(0)$ & 0.\\
			$i(0)$ & $3.6\times10^{-8}$.\\
			$r_{e}(0)$ & 0.\\
			$d_{e}(0)$ & 0.\\
			$v(0)$ &0.\\
			$v_{tot}(0)$ & 0.\\
			\hline
		\end{tabular}
		\label{int_val_tab}}
	\end{table}
	
	Tab.~\ref{int_val_tab} shows the initial values of the variables in our model. From this table we can say that $s(0)=1-i(0)\approx0.99$.
	
	\subsection{Without vaccination}
		In this subsection we will simulate to see various aspects of our model without vaccination. This implies that there exists only one important parameter which is the infection rate ($\beta (t)$).
		\begin{figure}[H]
			\centering
			\includegraphics[scale=0.34]{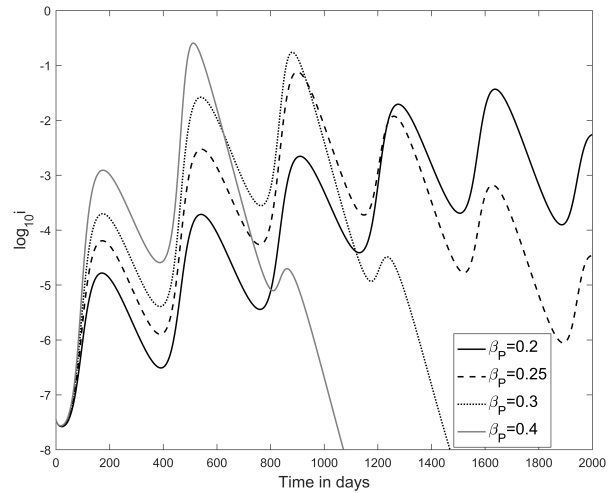}
			\caption{{Plot of $\log i$ vs time for different values of 
					$\beta_P$ (in $day^{-1}$).}\label{logi_nV}}	
		\end{figure}
		In Fig.~\ref{logi_nV} we have plotted the logarithmic value of the infectious cases with time for various peak values of the infection rate ($\beta (t)$). Here, we have assumed the frequency of the infection rate, $\omega=0.017 \:day^{-1}$. It is observed that for a low value of the peak of the infection rate ($\beta_{P}$) there are various peaks of the infectious cases of the disease and each peak is greater than the previous one. However, above a certain value of $\beta_{P}$, the second peak is the highest and the peaks that follow the second peak are lower.
		
		\begin{figure}[H]
			\begin{subfigure}{.49\textwidth}
				\centering
				\includegraphics[scale=0.26]{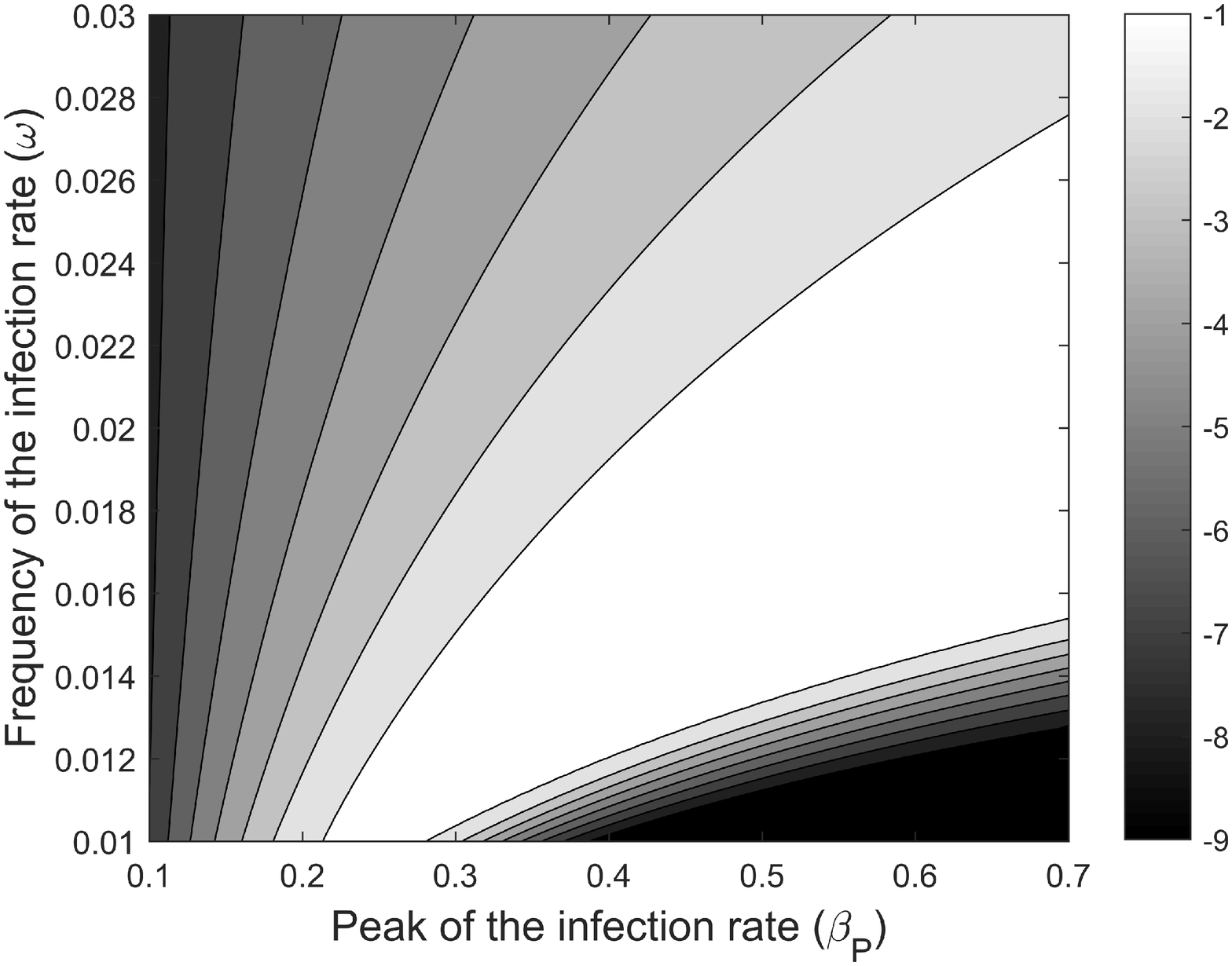}
				\caption{}
				\label{contour_nV_a}
			\end{subfigure}
			\begin{subfigure}{.49\textwidth}
				\centering
				\includegraphics[scale=0.26]{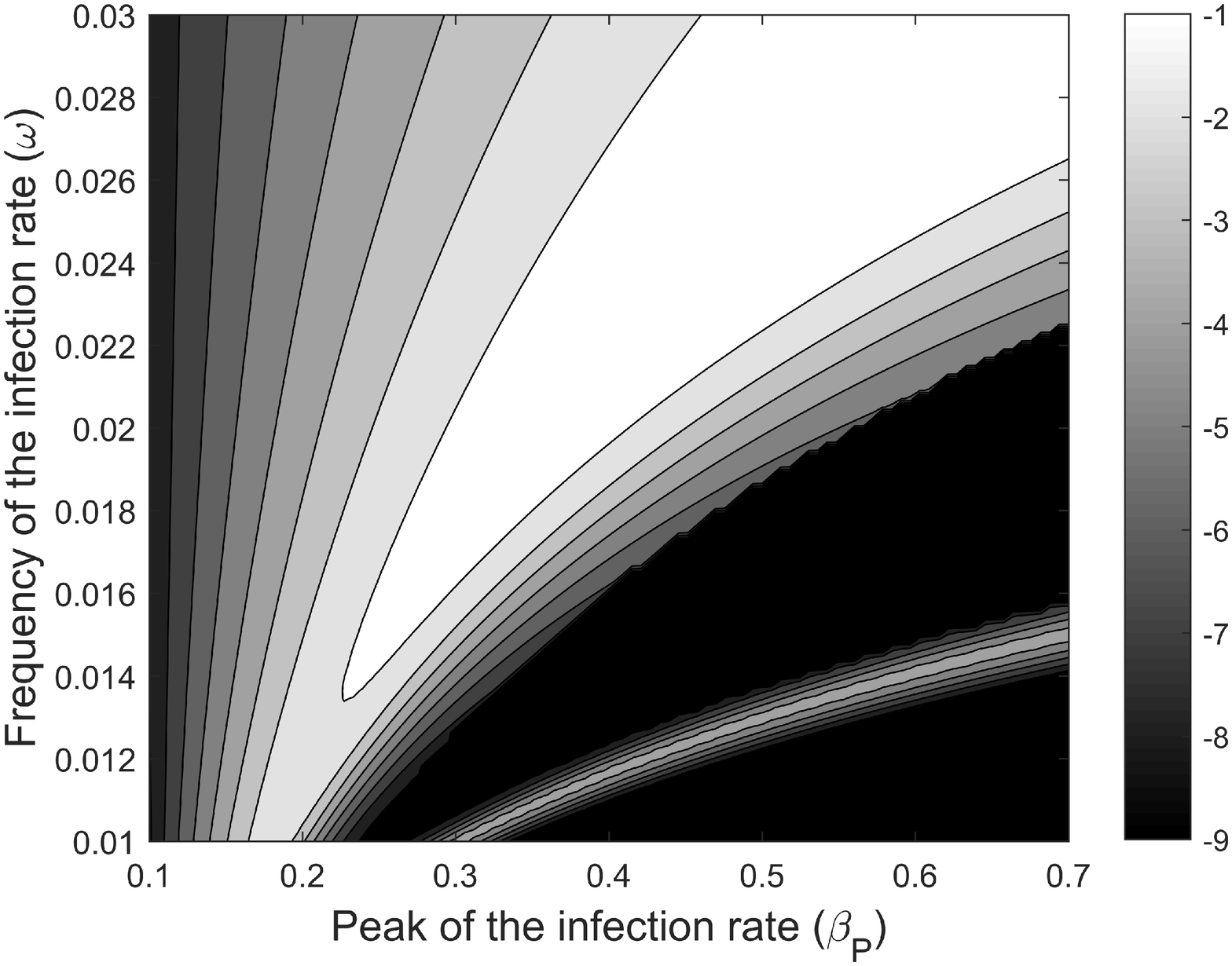}
				\caption{}
				\label{contour_nV_b}
			\end{subfigure}
			\caption{{Contour plots of the logarithmic values of second and third peak of the infectious cases for different values of $\beta_P$ and $\omega$. (a): contour plot of logarithmic values of the second peak. (b): contour plot of logarithmic values of the third peak.}\label{contour_nV}}	
		\end{figure}
	
		In Fig.~\ref{contour_nV} we have shown the contour plots of the logarithmic values of infectious cases ($\log i$) of the second and
		third peak for various values of the peak of the infection rate ($\beta_P$) and the frequency of the infection rate ($\omega$). From Fig.~\ref{contour_nV_a}, we can see that if $\omega$ decreases, the logarithmic value of the second peak 
		increases. However, from Fig.~\ref{contour_nV_b} it can be seen that the logarithmic value of the third peak approximately increases with $\omega$ (for a particular value $\beta_P$).
		
		The reason behind this can be well understood from the following argument. If the second peak is low, a large fraction of susceptible population is left to be infected. So, there is enough chance for the third peak to grow. Whereas, if the second peak is large, smaller fraction of susceptible people are left to be infected. Thus even if there might be a possibility of a third wave, the peak will be definitely small.
		
		In Fig.~\ref{contour_nV_b}, we can see that there is a region where $\log i\sim -9$. This region suggests that there is no third peak. It can also be found that when $\omega\lesssim 0.02$ $day^{-1}$, the number of infectious cases for the third peak are much lower compared to the second peak for our chosen range of $\beta_{P}$. Thus here we choose $\omega=\omega_{c}=0.02$ $day^{-1}$ as the cut-off value and all $\omega<\omega_{c}$ are ignored.
		\subsection{Vaccination included}
		In this subsection, we first show the variation of the logarithmic values of the fraction of the infectious cases as a function time, $t$ for a fixed value of the frequency of the infection rate, $\omega=0.02$ $day^{-1}$ after inclusion of the vaccination. Then we have studied the contour plot of the logarithmic values of the third peaks of the fraction of the infectious cases for different values of the peak of the infection rate ($\beta_{P}$) and the vaccination rate ($\lambda$). 
		
		Here, we have assumed that the vaccination is started at time $\tau$ after the initial time ($t=0$).

		\begin{figure}[H]
			\begin{subfigure}{.49\textwidth}
				\centering
				\includegraphics[scale=0.26]{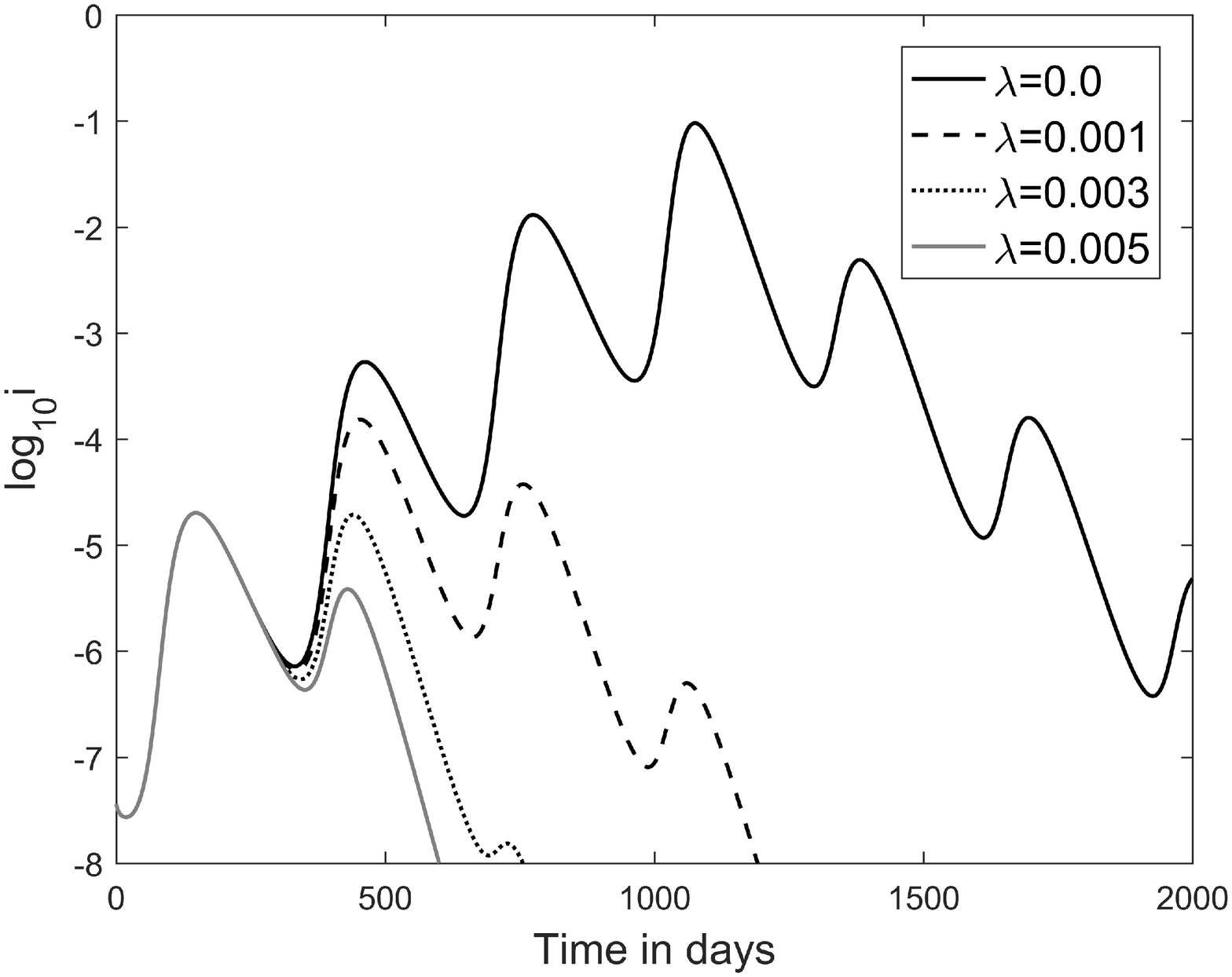}
				\caption{}
				\label{logi_wV_a}
			\end{subfigure}
			\begin{subfigure}{.49\textwidth}
				\centering
				\includegraphics[scale=0.26]{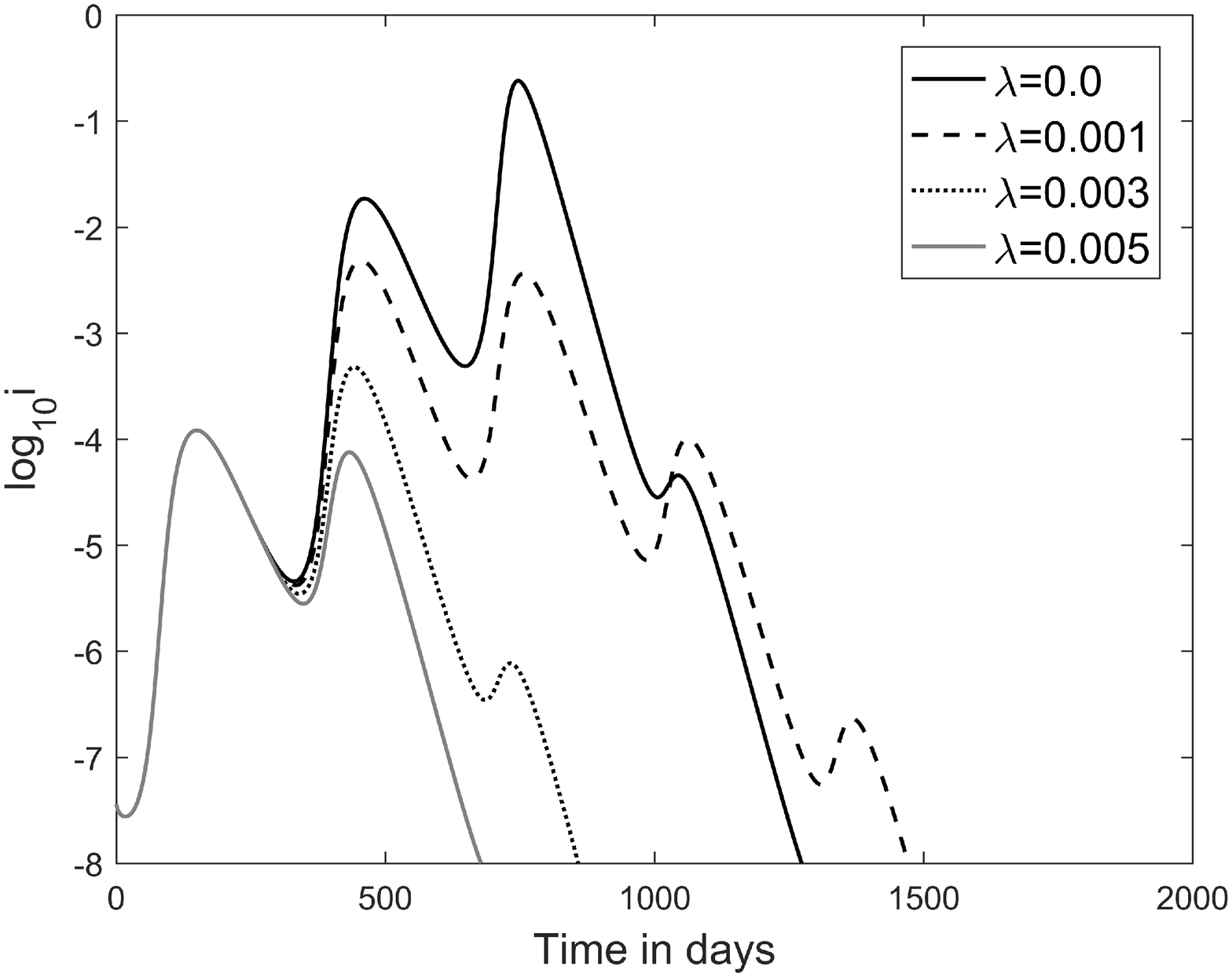}
				\caption{}
				\label{logi_wV_b}
			\end{subfigure}
			\caption{{Plot of $\log i$ vs time for different values of $\lambda$. (a): For 
				$\beta_P=0.25$ $day^{-1}$. (b): For $\beta_P=0.35$ $day^{-1}$.}\label{logi_wV}}	
		\end{figure}
		
		Fig.~\ref{logi_wV} represents the plots of the logarithmic values of the fraction of the infectious cases ($\log i$) as a function of time ($t$) for various values of the vaccination rate ($\lambda$). It is seen that for both plots, $\log i$ vs $t$ plot shows a decrement as the vaccination rate ($\lambda$) increase. Also, for both the plots it has been observed that above a value of the vaccination rate ($\lambda$), the third peak is suppressed. Thus the prediction of our model matches the expected behavior of the pandemic with the progression in vaccination. This phenomenon is observed by other authors also \cite{chaos_vac,covid_vac_india}. We also find that for lower values of the peak of the infection rate ($\beta_{P}$), the third peak can be controlled by a smaller vaccination rate ($\lambda$).
		
		Now for this particular value of $\omega$, we want to see how the logarithmic values of third peak of the fraction of the infectious cases varies for different values of $\beta_P$ and $\lambda$.  
		\begin{figure}[H]
			\begin{subfigure}{.49\textwidth}
				\centering
				\includegraphics[scale=0.30]{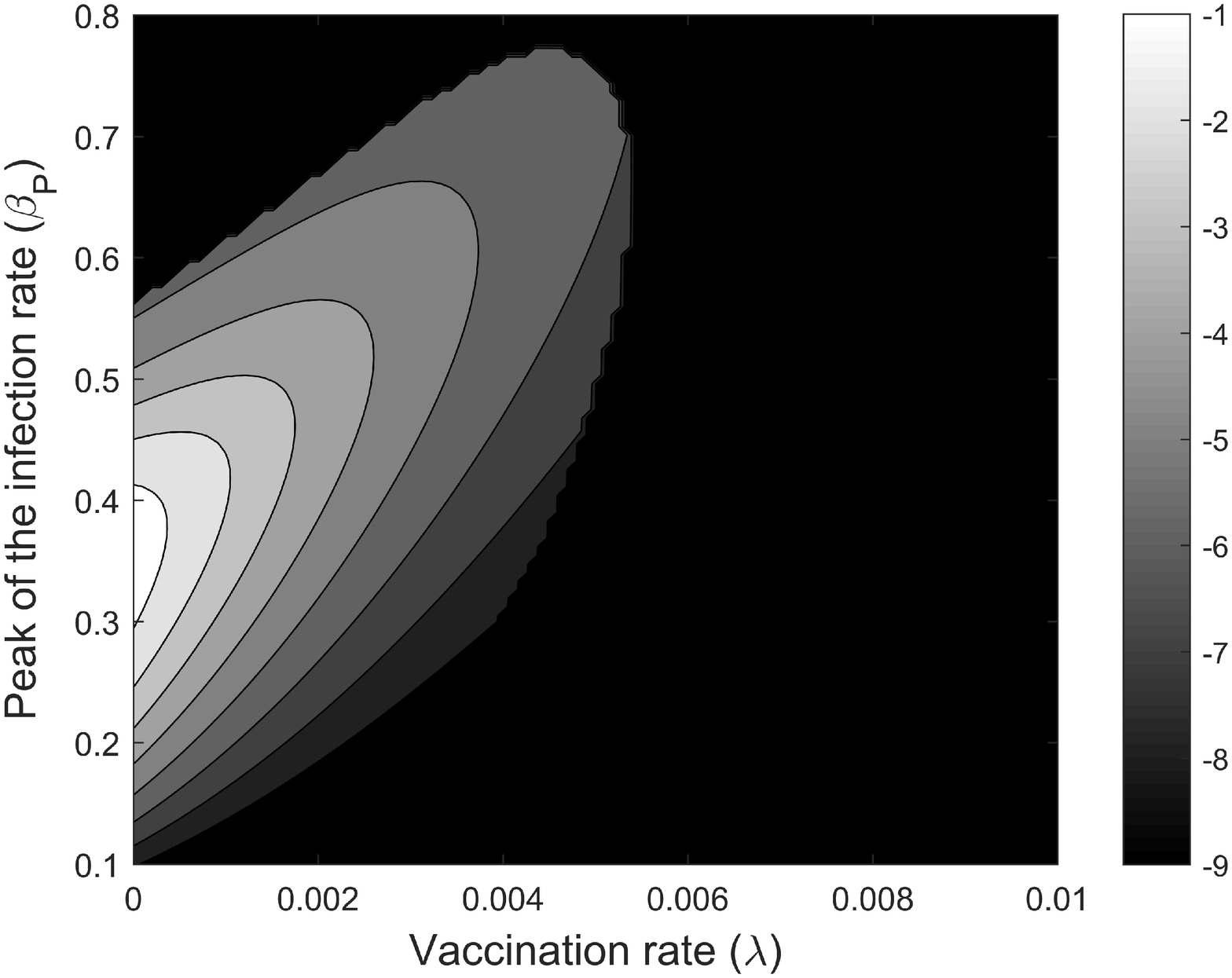}
				\caption{}
				\label{contour_wV_a}
			\end{subfigure}
			\begin{subfigure}{.49\textwidth}
				\centering
				\includegraphics[scale=0.30]{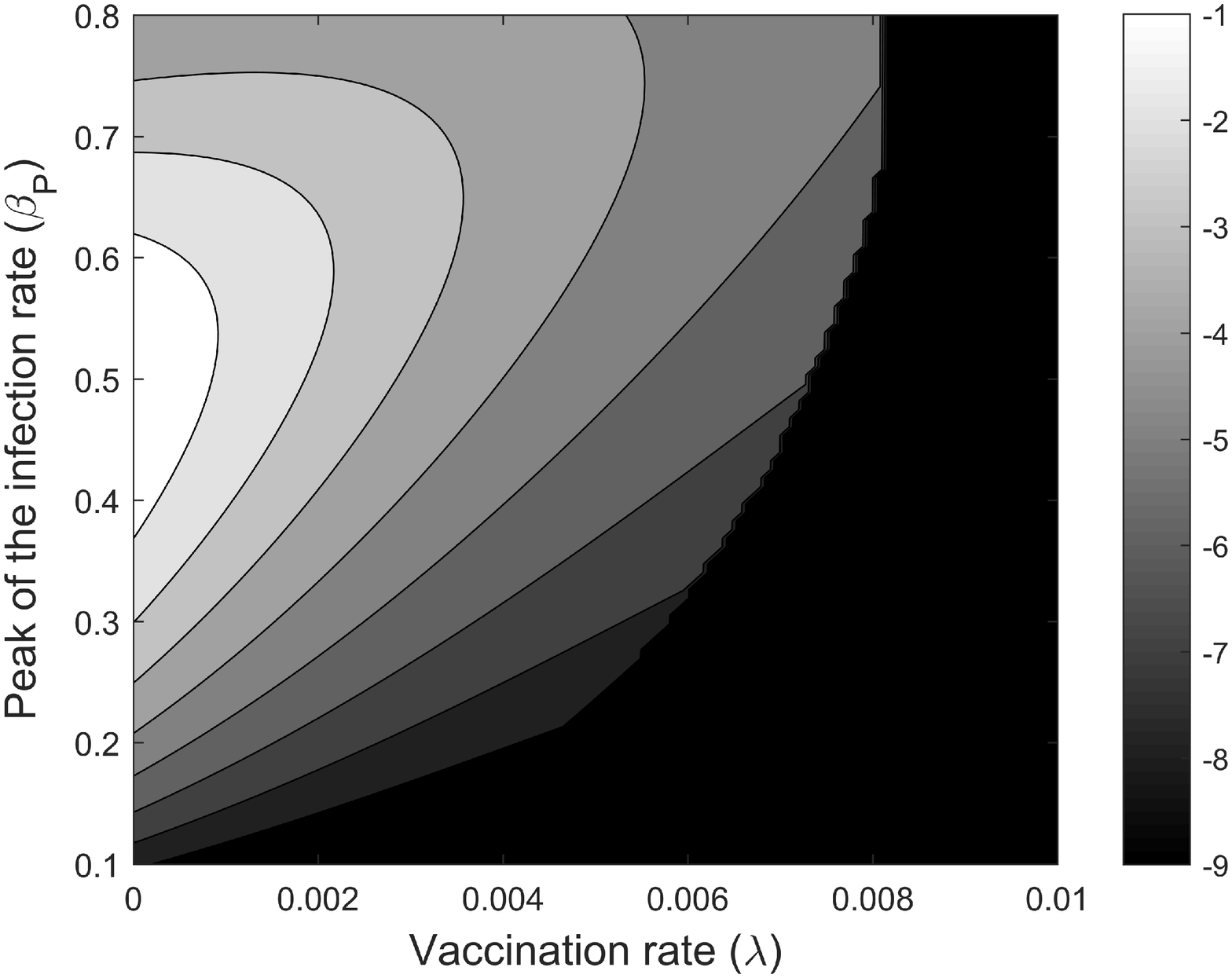}
				\caption{}
				\label{contour_wV_b}
			\end{subfigure}
			\caption{{Contour plots of the logarithmic values of third peak of 
					the infectious cases for $\omega=0.02$ $day^{-1}$ and $\omega=0.025$ $day^{-1}$ with respect to 
					$\beta_P$ and $\lambda$. (a): contour plot of logarithmic values of the third 
					peak for $\omega=0.02$ $day^{-1}$. (b): contour plot of logarithmic values of the third 
					peak for $\omega=0.025$ $day^{-1}$.}\label{contour_wV}}	
		\end{figure}

		Fig.~\ref{contour_wV_a} shows that if $\lambda\approx0.005$ $day^{-1}$ then the third peak will not arise whatever be the value of $\beta_P$. We can consider it as a cut-off value of the vaccination rate ($\lambda_{c}$). This implies, minimum of 63.2\% of the population have to be vaccinated in $\sim200$ days from the start of the vaccination to prevent the recurrence of the this pandemic in the form of a third wave. Also, from Fig.~\ref{contour_wV_a} we can say that if $\lambda\approx0.0025$ $day^{-1}$, large number of infectious cases in third wave can be avoided. This implies a minimum of 63.2\% of the population has to be vaccinated in $\sim400$ days from the start of vaccination to prevent a large third wave.
		
		\begin{figure}[H]
			\centering
			\includegraphics[scale=0.34]{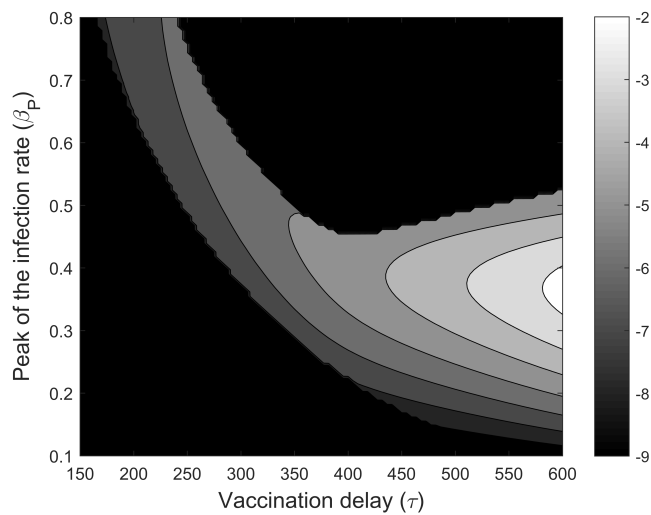}
			\caption{{Contour plots of the logarithmic values of third peak of 
					the infectious cases for different values of $\beta_P$ and $\tau$.}\label{contour_wV_tau}}	
		\end{figure}
	
		In Fig.~\ref{contour_wV_tau}, we have shown the contour plot of $\log i$ values of the third peak with respect to peak of the infection rate ($\beta_P$) and vaccination delay ($\tau$) for the vaccination rate $\lambda=0.002$ $day^{-1}$. It can be seen that infectious cases in the third wave increase with increasing vaccination delay ($\tau$). This implies that besides the vaccination rate ($\lambda$), it is important to start vaccination as early as possible to counter the third wave. 
		
		We would also like to see the variation for the third peak for $\omega>0.02$ $day^{-1}$. In Fig.~\ref{contour_wV_b} we have drawn the contour plot for $\omega=0.025$ $day^{-1}$. Here, it is seen that when $\omega$ increases, the contours get distributed out as a function of the vaccination rate ($\lambda$). So, we can conclude that whatever be the value of $\beta_P$, the vaccination rate ($\lambda$) should be increased for an increased value of $\omega$. This implies that the cut-off value of the vaccination rate ($\lambda_{c}$) increases that the vallue which is discussed earlier in this section,with increasing $\omega$. In other words, if the period of the infection rate ($T$) is decreased, 63.2\% of the population should be vaccinated at a faster rate than which is discussed to avoid the third wave of the COVID-19 pandemic.
		
		Thus, it can be concluded that our robust and minimal model encompasses several different realistic aspects of the pandemic and can be used to make predictions about the various future possibilities suitably. 
\section{Conclusions}
	In this section, we will summarize the important features and predictions of our 
	model. Here we assume the SEIRD (Susceptible-Exposed-Infectious-Recovered-Dead) model and modify it by 
	introducing the vaccination term. 
	
	The main assumptions of our model are enlisted below:
	\begin{itemize}
		\item We have assumed that vaccination started with a delay $\tau$ from the initial time ($t=0$).
		
		\item It has been considered that even after vaccination a person can be infected by COVID-19 within a fixed period which is represented by $\frac{1}{k}$. This can be interpreted as the time a person is required to develop immunity after vaccination.
		
		\item We have assumed a particular cyclical form of the infection rate ($\beta (t)$) as a function of time $t$. The infection rate ($\beta (t)$) is one of the most important parameters of our study.
		
		\item Other most important parameter, other than the infection rate ($\beta (t)$),  is the vaccination rate ($\lambda$). Our main focus is to study the interplay between these two when other parameters of the model are fixed.
		
		\item All the parameters other than $\beta(t)$ and $\lambda$ are argued to have minimal effect on the dynamics of the model. Thus some fixed values have been used for the entire study which is suitable for COVID-19.
	\end{itemize}	
	Next, we move on to the predictions of our model:
	\begin{itemize}
		\item In section 3, it has been shown that under certain approximations, the inverse of the vaccination rate ($\lambda$) represents the time to vaccinate $(1-\frac{1}{e})=0.632$ fraction or 63.2\% of the population. It is worth noting, this fraction of the population has to be vaccinated within a certain period to avoid the possibility of a third wave as has been discussed in section~\ref{sim_section}.
 
		\item We have found that there can be many waves for this pandemic for a small value of the peak of the infection rate ($\beta_{P}$) without vaccination. There will be fewer waves if $\beta_{P}$ is high, since most of the population will already be infected by the second or third wave. For high value of $\beta_{P}$, the peaks of the fraction of the infectious population are relatively higher for the second and third wave as compared to lower values of $\beta_{P}$ (Refer to Fig.~\ref{contour_nV} for details). 
		
		\item From Fig.~\ref{contour_nV_b} we can conclude that for any frequency of the infection rate ($\omega$) which lie below $\omega\approx0.02$ $day^{-1}$, the third peak is quite insignificant any $\beta_P$. Thus we have assumed $\omega=0.02$ $day^{-1}$ as the lower cut-off value for all cases with or without vaccination.
		
		\item For the simulation in which vaccination has been included, $\omega$ has been set to a value $0.02$ $day^{-1}$. We have taken two different values of $\beta_{P}$. We have found that the peak values of the infectious population decrease as the vaccination rate ($\lambda$) increases. Also, for lower values of $\beta_{P}$, lower values of the vaccination rate are sufficient to suppress the third peak. 
		
		\item From Fig.~\ref{contour_wV_a} we conclude that 63.2\% of the population has to be necessarily vaccinated within a period of $\sim200$ days from the starting of the vaccination time ($\tau$) to suppress the third wave of COVID-19.
		
		\item From our analysis we can also conclude that if 63.2\% of the population is vaccinated within a period of $\sim 400$ $days$ from the start of the vaccination ($\tau$), it might still be possible to avoid a large third wave of COVID-19 pandemic. 
		
		\item Also the onset time of vaccination has a non-trivial effect on $\lambda_{c}$. We observed that if the vaccination is delayed, the vaccination rate ($\lambda$) has to be increased further to achieve vaccination of 63.2\% of the population in a short time to ensure a minimal third wave of pandemic.
		
		\item From Fig.~\ref{contour_wV_b} it is seen that as the frequency of the infection rate ($\omega$) increases the contours of the logarithmic values of the fraction of infectious population ($\log i$) are distributed out as a function of the vaccination rate. Thus if $\omega$ increases, the cut-off value of the vaccination rate should also be increased to avoid any third wave. 
	\end{itemize}

	Finally, we would like to conclude that these predictions are governed by the assumptions that have gone into the model. In reality, the situation can be far more complex since it involves various aspects of human and socio-economic factors as also the behavior of the virus itself in terms of its different mutations and its virulence. So, the infection rate ($\beta(t)$) can be deduced to be nonlinear and complicated. Here, we also note that the requirement of booster doses of the vaccine after a period of 6 to 8 months is already being indicated by the medical fraternity to boost the level of immunity against this virus for a sustained period of time, particularly among aged people with multiple co-morbidities. In future, we hope to study these complex aspects to model the behavior of pandemics or epidemics more effectively.
	
	\section*{Acknowledgment}
	
	The authors would like to thank Dr. Indrani Bose, Dr. Tapati Dutta, and Dr. Sujata Tarafdar for their valuable and useful comments and suggestions. The authors would also like to thank the Department of Physics, St. Xavier's College, Kolkata for providing support during this work. One of the authors (S. C.) acknowledges the financial support provided from the University Grant Commission (UGC) of the Government of India, in the form of CSIR-UGC NET-JRF. Finally, the authors would also like to express their gratitude to the anonymous referee for his/her valuable comments and suggestions.
	
	\bibliographystyle{ws-ijmpc}
	\nocite{*}
	\bibliography{References}	
\end{document}